\documentclass[layout=twocolumn]{achemso}

\usepackage{graphicx}
\usepackage{amsmath}
\usepackage{amssymb}
\usepackage[utf8]{inputenc}
\usepackage{dcolumn}
\usepackage{bm}
\usepackage{psfrag}
\usepackage{yfonts}
\usepackage{version}
\usepackage{natbib}
\usepackage{units}
\usepackage{epsfig}
\usepackage[usenames,dvipsnames]{color}

\newcommand{\cm}{\ {\rm cm}^{-1}}

\usepackage{ulem}  
\newlength{\mylenunit}
\title{Suppression of Quantum Oscillations and Dependence on Site Energies in Electronic Excitation Transfer in the Fenna-Matthews-Olson Trimer }

\author{Gerhard Ritschel}
\affiliation{Max-Planck-Institut f\"ur Physik komplexer Systeme, N\"othnitzer Str.~38, D-01187 Dresden, Germany}

\author{Jan Roden}
\affiliation{Max-Planck-Institut f\"ur Physik komplexer Systeme, N\"othnitzer Str.~38, D-01187 Dresden, Germany}

\author{Walter T.\ Strunz}
\affiliation{Institut f\"{u}r Theoretische Physik, Technische Universit\"at Dresden, D-01062 Dresden, Germany}

\author{Al\'an Aspuru-Guzik}
\affiliation{Department of Chemistry and Chemical Biology, Harvard University, 12 Oxford Street, Cambridge, MA 02138}

\author{Alexander Eisfeld}
\email{eisfeld@mpipks-dresden.mpg.de}
\affiliation{Department of Chemistry and Chemical Biology, Harvard University, 12 Oxford Street, Cambridge, MA 02138}
\alsoaffiliation{Max-Planck-Institut f\"ur Physik komplexer Systeme, N\"othnitzer Str.~38, D-01187 Dresden, Germany}

\date{\today}


\begin{document}

\begin{abstract}
Energy transfer in the photosynthetic Fenna-Matthews-Olson (FMO) complex of Green Sulfur Bacteria is studied numerically taking all three subunits (monomers) of the FMO trimer and the recently found eighth bacteriochlorophyll (BChl) molecule into account.
The coupling to the non-Markovian environment is treated with a master equation derived from non-Markovian quantum state diffusion.
When the excited state dynamics is initialized at site eight, which is believed to play an important role in receiving excitation from the main light harvesting antenna, we see a slow exponential-like decay of the excitation.
This is in contrast to the oscillations and a relatively fast transfer that usually occurs when initialization at sites 1 or 6 is considered.
We show that different values of the electronic transition energies found in the literature can lead to large differences in the transfer dynamics and may cause additional suppression or enhancement of oscillations.

{\bf TOC Graphic.}

\begin{figure}
\includegraphics[width=2in]{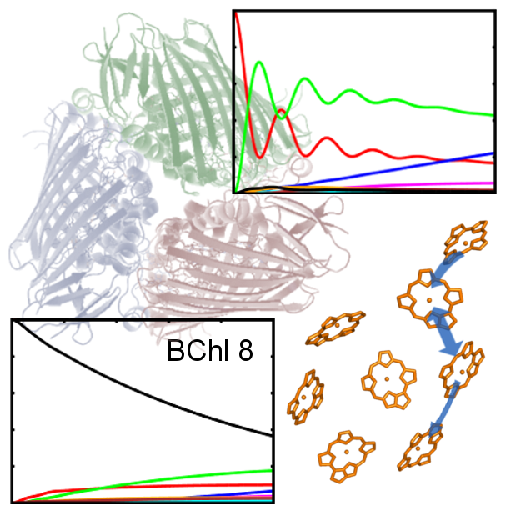}
\end{figure}


\end{abstract}


\newpage
\begin{figure}
\includegraphics[width=0.7\mylenunit]{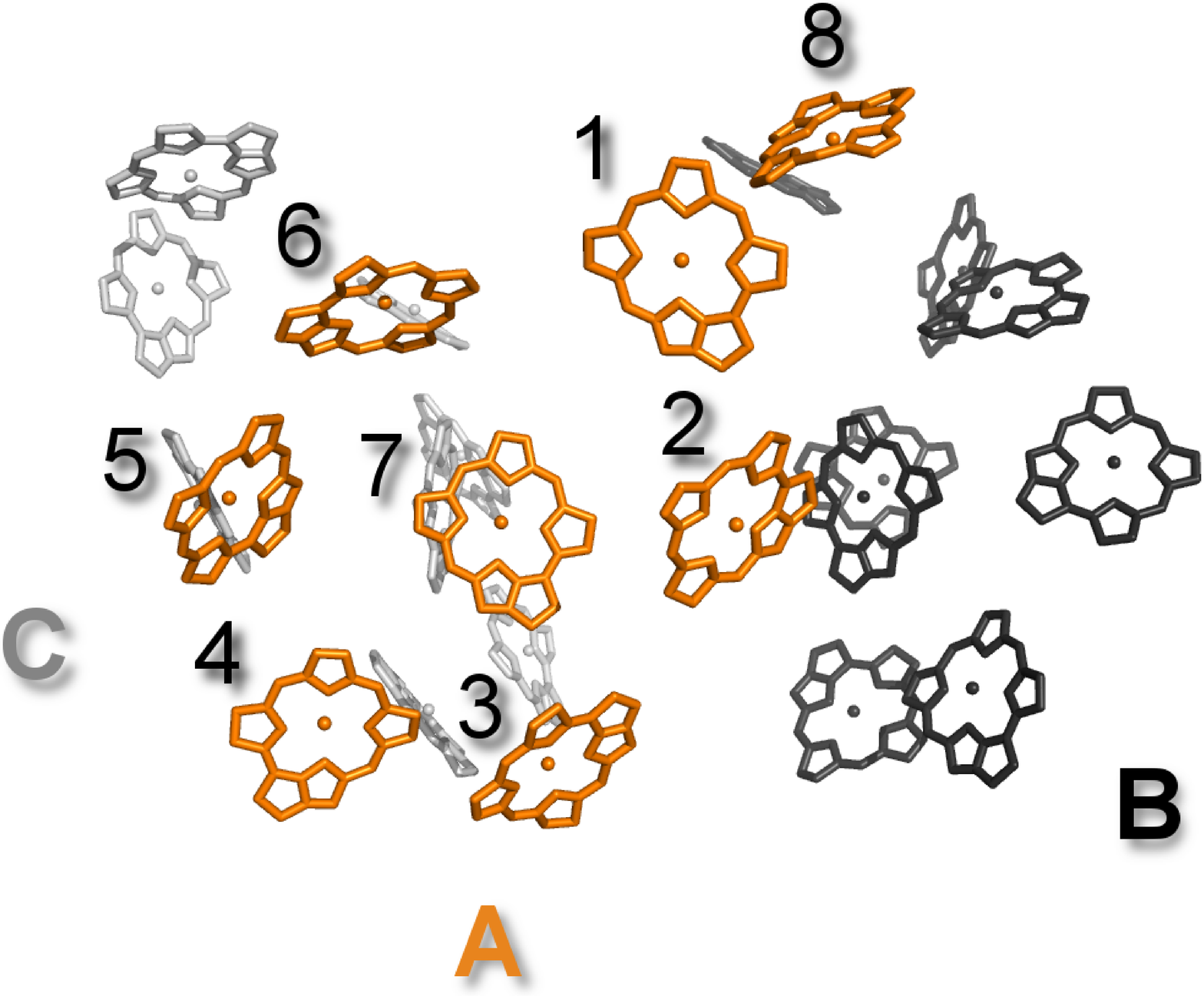}
\hspace{0.1\mylenunit}
\includegraphics[width=0.6\mylenunit]{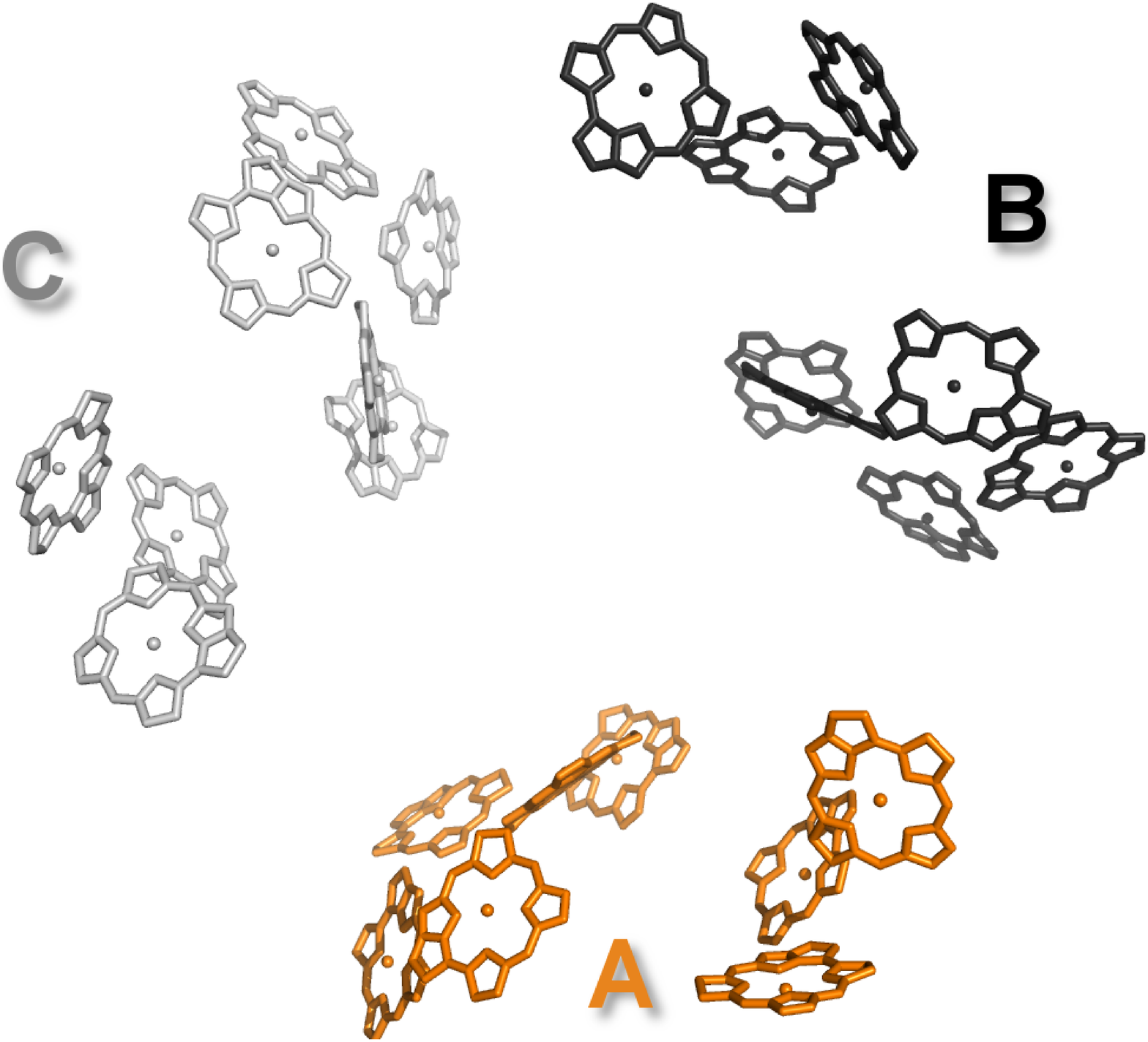}
\caption{
The Fenna-Matthews-Olson photosynthetic complex trimer, consisting of symmetry-equivalent monomers A, B, and C.
Top: side view and labeling of the BChl molecules within one monomer.
Bottom: top view.
The figure was created using PyMOL~\cite{PyMOL} and is based on the PDB entry 3ENI.
}
\label{fig:FMO}
\end{figure}

The Fenna-Matthews-Olson (FMO) complex bridges electronic excitation energy transfer from the chlorosome (the main light harvesting antenna complex) to the photosynthetic reaction center in green sulfur bacteria~\cite{ScMuEl11_93_,AmVaGr00__}.
It consists of three identical subunits, which we denote as  monomers A, B, and C (see \ref{fig:FMO}).
Each of the monomers contains eight BChl molecules~\cite{BeFrNe04_274_,TrWeGa09_79_}.
Electronic excitation can be transferred between the BChl molecules via transition dipole-dipole interaction.
Due to the availability of high resolution crystal structure a good overall understanding of the transition energies of the chlorophyll molecules and their dipole-dipole couplings has been obtained, although there is still a large variance between values found in the literature  (see Ref.~\cite{OlJaLi11_8609_} and in particular Ref.~\cite{MiBrGr10_257_}).

Regarding the arrangement of the FMO complex w.r.t.\ the chlorosomes, it is assumed that BChls~1, 6, and~8 are located at the baseplate (which connects the chlorosomes and the FMO complex) and receive electronic excitation captured by the chlorosomes.
In Ref.~\cite{ScMuEl11_93_} evidence was found that the newly discovered eighth BChl directly obtains a significant part of the excitation handed over to the FMO complex. 
BChls 3 and 4 are located in the vicinity of the reaction center.
This arrangement of the BChls is in accordance with the ordering of their electronic transition energies.
BChl 8,  BChl 1 and 6 which are located near the baseplate have large transition energies.
Thus the FMO complex acts like a funnel that transfers energy from one side to the other, where BChls 3 and 4 have low transition energies.
Since the eighth BChl molecule was discovered only recently, it has not been considered in most of the previous studies (e.g.\ Refs.~\cite{AmVaGr00__,ReMaKue01_137_,ReMa98_4381_,MiBrGr10_257_,ChFl09_241_,BrStVa05_625_,EnCaRe07_782_,CaChDa09_105106_,PlHu08_113019_,MueMaAd07_16862_,MoReLl08_174106_,AdRe06_2778_,KrKrRo0__,WuLiSh10_105012_,SaChWh11_11906_,BrEi11_051911_}).
Only in some recent studies the energy transfer has been treated taking the eighth BChl molecule into account~\cite{OlJaLi11_8609_,ScMuEl11_93_}.

In the present work we investigate theoretically excitation transfer in the full FMO trimer, focusing in particular on the role of BChl 8 and the dependence of the transfer on different sets of transition energies.
For our calculations we use a master equation which is derived from the non-Markovian quantum state diffusion (NMQSD) equation~\cite{DiSt97_569_,DiGiSt98_1699_,YuDiGi99_91_,StYu04_052115_,Yu04_062107_,VeAlGa05_124106_}, within the zeroth order functional expansion (ZOFE) approximation~\cite{YuDiGi99_91_,RoEiWo09_058301_,RoStEi11_034902_}. 
In a recent publication we have shown  (for the case of a single monomer subunit, taking only seven BChl molecules into account) that with this approach we obtain very good agreement with numerically exact results~\cite{MONOMER}. 
This method, described in Ref.~\cite{MONOMER}, allows one to calculate the energy transfer in the full FMO trimer consisting of 24 coupled BChls within a few hours on a standard PC, taking  coupling of the electronic excitation to the non-Markovian environment into account.  
Our master equation is derived from a stochastic Schr\"{o}dinger equation by averaging over many trajectories \cite{RoEiWo09_058301_,StYu04_052115_}.
For individual trajectories the coherence time might be substantially longer than for the average.
Similar conclusions have been drawn recently \cite{IsFl11_6227_}. However, note, that the interpretation of such individual trajectories is far from trivial (see e.g.\ \cite{Di08_080401_,WiGa08_080401_}).

\label{sec:full_trimer}
 
\begin{figure}
\includegraphics[width=0.8\mylenunit]{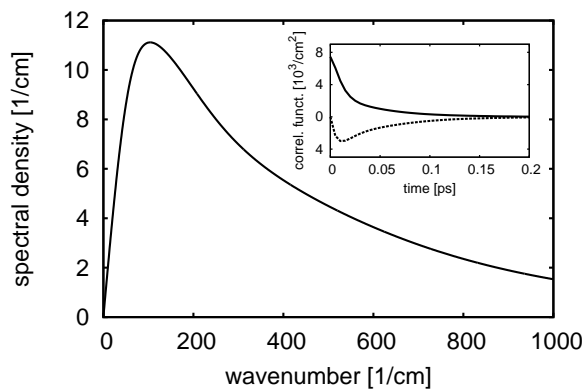}
\caption{\label{fig:spd}
The spectral density employed, taken from Ref.~\cite{MONOMER}.
The reorganization energy of the spectral density is 35 $\cm$, which is in the order of the interactions between the BChls and leads to a decoherence time of roughly 0.15 ps.
The environment correlation function corresponding to this spectral density is shown as inset (dashed line: imaginary part, solid line: real part). 
The correlation time is much shorter than the time range on which we consider the energy transfer.}
\end{figure}

Although the method enables us to treat complicated structured spectral densities for the exciton-vibrational coupling, in the present paper we use a relatively simple spectral density shown in \ref{fig:spd}, which nevertheless leads basically to the same non-Markovian dynamics as the spectral density used in  Ref.~\cite{IsFl09_17255_}, as was shown in  Ref.~\cite{MONOMER}.

Important parameters are the interactions between the chromophores and the site energies. 
For the interaction matrix elements between the BChl molecules we take values from Ref.~\cite{OlJaLi11_8609_}, which are given in \ref{tab:coup_monomer} for the interactions within one monomer and in \ref{tab:coup_trimer} for the interactions between the BChl molecules located in different monomers.
Note that these values and the values of the site energies used in the following are only according to a working model developed using biochemical and spectroscopic evidence.

Following Ref.~\cite{ScMuEl11_93_}, we assign BChl~8 to the monomers such that the strongly interacting neighbors BChl~1 and BChl~8 belong to the same monomer (see \ref{fig:FMO}) and the coupling between different monomers is relatively weak ($< 8\cm$, see \ref{tab:coup_trimer}). Note that this assignment is different from the one of Ref.~\cite{OlJaLi11_8609_}, where BChl~8 is assigned to the monomers based on the protein structure.

For the site energies  of the BChls we consider values taken from Olbrich et.\ al.\ (OLB) in Ref.~\cite{OlJaLi11_8609_} and also values of Schmidt am Busch et.\ al.\ (SAB) in Ref.~\cite{ScMuEl11_93_}, which are both shown here in \ref{tab:site_energies}.
Note the large differences between the two sets of site energies. 
These two sets of transition energies for the BChls stem from recent publications and are based on the crystal structure.
However, quite different strategies have been employed to obtain them.
In Ref.~\cite{ScMuEl11_93_} (SAB) quantum chemistry on BChl ground and excited states is combined with electrostatic calculations, where the whole protein is included in a very detailed fashion. 
In contrast, in Ref.~\cite{OlJaLi11_8609_} (OLB) a time-dependent approach has been chosen in which many molecular dynamics ground state trajectories have been run, calculating the electronic gap energy along the trajectory. This does not lead to a single transition energy but to a distribution for each BChl.
This distribution was found to be asymmetric with tails towards higher energies and different for each BChl.
In the present work we have taken the average positions of these distributions to represent the electronic transition energy.

Although there exist many sets of empirical parameters in the literature (for a discussion see Ref.~\cite{MiBrGr10_257_}), we have chosen these two sets of values, since they are both obtained from a model taking the eighth BChl molecule into account.

\begin{figure}
\includegraphics[width=\columnwidth]{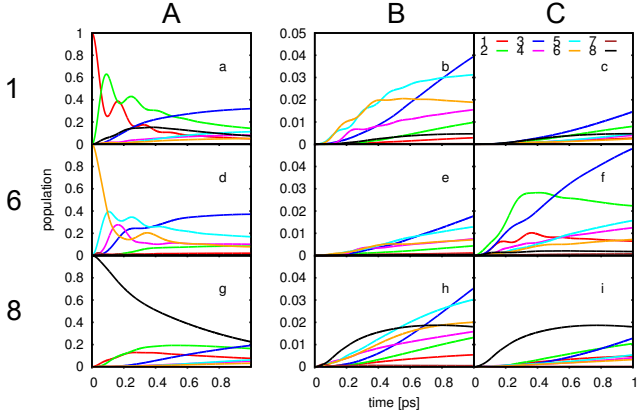}
\caption{\label{fig:transfer77K} Calculated transfer through the FMO trimer at $T=77\,\rm{K}$ for the energies of Ref.~\cite{OlJaLi11_8609_} (OLB).
From left to right: Population on the BChls of monomer subunits A, B, C.
First row: initial excitation on BChl~A1.
Second row: on BChl~A6.
Third row: on BChl~A8.
Note that the populations on monomer B and C are on a different scale than those for monomer A, because of the very small population that is transferred.}
\end{figure}

\begin{figure}[t]
\includegraphics[width=\columnwidth]{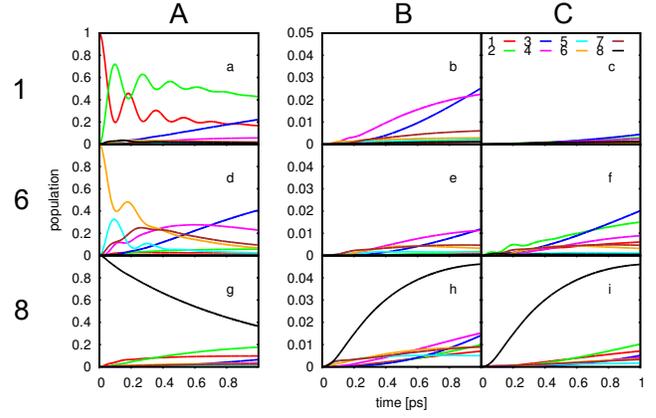}
\caption{\label{fig:transfer77K_SchmidtaB_energies} As \ref{fig:transfer77K}, but with the SAB energies taken from Ref.~\cite{ScMuEl11_93_} (see \ref{tab:site_energies}). 
}
\end{figure}

The calculated energy transfer in the FMO for the site energies of Ref.~\cite{OlJaLi11_8609_} is shown in \ref{fig:transfer77K} and for those of Ref.~\cite{ScMuEl11_93_} in  \ref{fig:transfer77K_SchmidtaB_energies}.
The results presented here are for a low temperature\footnote{The values in OLB are based on a room temperature MD. However, we do not expect a strong temperature dependence for the average values.} of  77 K for three different initial conditions.
In the first row all excitation is located initially on BChl~1 of monomer~A.
In the second row it is localized on BChl~6 and in the third row on  BChl~8 \footnote{Due to the rotational symmetry of the FMO trimer, this is equivalent to the case where the excitation would reside on monomer B or C.}. 

The first thing to note is that for both sets of site energies and all three initial conditions there is only little transfer away from monomer A, which is due to the very small inter-monomer interactions (see \ref{tab:coup_trimer}). 
We have also investigated initial conditions for which the excitation was delocalized on the three monomers and found, as expected, essentially the same behavior.

We now take a closer look at the different initial conditions.
We first treat the transport along the FMO for the site energies of OLB  and contrast them later to the transfer dynamics obtained for the site energies of SAB.

One sees that for the initial excitation on BChl~1 there are strong oscillations of the excitation between BChls~1 and~2.
This is due to the strong coupling between these two BChls (see \ref{tab:coup_monomer}).
Similar dynamics, but with much weaker oscillations occur when the excitation is initialized on BChl~6 (second row of \ref{fig:transfer77K}).

Different  dynamics is found for the site energies of SAB shown in \ref{fig:transfer77K_SchmidtaB_energies}.
When starting on BChl 1 the oscillations between BChl 1 and 2 are even more pronounced and last longer than for the parameters of OLB. In the second row (starting on BChl 6) a stronger oscillation between BChl 6 and 5 is now visible. 
A large difference in the transport between the two sets of site energies becomes apparent when considering the time dependence of the populations of BChl 3, which has the lowest site energy and thus in thermal equilibrium is expected to have the highest population.
While for the OLB energies there is already a saturation after about 0.6~ps , for the SAB site energies the populations on BChl 3 are still growing at 1.0 ps. 
This strong difference can be attributed to the much lower energy of BChl 3 w.r.t.\ the energies of the other BChls for the SAB site energies compared to the OLB site energies.
Note further that for the OLB site energies BChl~7 is detuned strongly from all other site energies and is thus excluded from the transfer. The effective spread of the site energies, without BChl~7, is then only $\sim 220 \cm$ and thus much smaller than in the SAB case, where it is $\sim 500 \cm$. From this, one expects that the transfer is faster in the OLB case, in agreement with our observation.

Remarkably, a completely different dynamics takes place when the eighth BChl is excited initially (third row): 
In this case the excitation on BChl~8 decays very slowly with an exponential-like curve and no oscillations are present.
This suppression of oscillations (see also Refs~\cite{ReRaMu11_075102_,ARXIV_Moix}) is due to the fact that the only relevant coupling of BChl~8 is to BChl~1 and that the population that is handed over to BChl~1 is immediately transferred to BChl~2 due to the large coupling between BChl~1 and~2.
Thus, no population can be accumulated on BChl~1 to initialize oscillations between BChl~1 and~2.
A simple model is given in Ref.~\cite{ARXIV_Moix}, using the SAB site energies.
In Ref.~\cite{ARXIV_Moix} the slow exponential-like decay is explained by the fact that the coupling of BChl~8 to BChl~1 is small compared to the energy gap between BChl~8 and~1 and the coupling to the environment.
Indeed, for the SAB site energies the energy gap is $~200 \cm$ and thus much larger than the coupling of $~20 \cm$ between BChl~8 and~1. However, for the OLB energies the gap is only $~30 \cm$, i.e.\ comparable to the coupling between BChl~8 and~1.
We have found that for short times ($< 0.1$~ps) the dynamics is dominated by the (slow) coherent oscillation between BChl 8 and the strongly coupled dimer consisting of BChl 1 and 2.  For later times decoherence leads to the observed monotonic decay of the population.

We also considered the transfer at a higher temperature of 300~K and found that it  is similar to that at 77~K shown here, but the oscillations are less pronounced.
When BChl~8 is excited initially, the transfer at 77~K and 300~K looks nearly the same.

Furthermore, we also calculated the transfer for the energies and couplings of Ref.~\cite{OlJaLi11_8609_}, obtained from the crystal structure instead of the MD simulations, and found almost identical transfer dynamics.

Regarding the effects of the variations of the site energies, one should keep in mind that the values of OLB stem from an atomistic model based on molecular dynamics simulations, which yield distributions for the site energies \cite{OlJaLi11_8609_,ARXIV_Sangwoo}, which  in  the case of OLB show a pronounced asymmetry.
In the present work we have used the mean values of these distributions as the site energies. We have also performed calculations using the maxima of the distributions and found qualitatively the same transfer dynamics as for the mean values.  

Beside the electronic transition energies and couplings also the interaction with vibrational modes plays an important role.
While in our previous work we have investigated structured spectral densities \cite{RoEiWo09_058301_,RoStEi10_5060_,MONOMER}, here we have taken an often used featureless  model spectral density.
From atomistic simulations it is also possible to deduce spectral densities \cite{OlStSc11_1771_,ARXIV_Sangwoo}.
While the one obtained in \cite{ARXIV_Sangwoo} has a much smaller effective
reorganization energy (a measure for the strength of the coupling to the
environment)\footnote{In Ref.~\cite{MONOMER} we defined the effective
reorganization energy of the environment spectral density to quantify the
part of the overall reorganization energy that is in the relevant frequency
region, where the purely electronic dynamics couple to the environment.
For the spectral density of \ref{fig:spd} the effective reorganization
energy is roughly 30$\cm$.} than the spectral density used in the present work,
the spectral density of OLB \cite{OlStSc11_1771_} has an effective
reorganization energy which is an order of magnitude larger. 
We have performed calculations for different effective reorganization energies by  scaling the whole spectral density with a certain factor.
 As expected, we found that with decreasing/increasing reorganization energy the observed beatings become stronger/weaker. 
For an increase of the original spectral density of \ref{fig:spd} by a factor of four  (then the increased spectral density has roughly the same effective reorganization energy as the averaged spectral density for BChl 1--6 obtained in Ref.~\cite{OlStSc11_1771_}), we observed that the beatings have vanished almost completely. 
Remarkably, for this increased spectral density we found the population on BChl~3 after 1~ps to be about three times larger than the population for the original spectral density.
This shows the strong influence of the coupling to the environment on the transfer dynamics of the FMO complex.
We have also performed calculations for a Markovian environment. Compared to the spectral density used in the present work, we found in the Markovian case that the population that arrives on BChl 3 after 1~ps is at least by roughly a factor of two lower -- regardless of the coupling strength to the Markovian environment.

\label{sec:summary}

In the present paper we considered the energy transfer through the full FMO trimer, where we also included the eighth BChl molecule.
The values for the energies and couplings between the BChl molecules were taken from recent MD simulations and from the crystal structure (from Refs.~\cite{OlJaLi11_8609_,ScMuEl11_93_}). 
For the calculation we used an efficient non-Markovian master equation (derived from the NMQSD).

It was found, as expected, that due to the small couplings between the BChl molecules of different monomers (compared to the chromophore-protein coupling), there is only little exchange of electronic excitation between the monomer units (this weak coupling limit has been discussed e.g. in Ref.~\cite{AmVaGr00__}).
This means that the three monomers act as independent transfer channels.
Thus, the limitation to only a single isolated monomer unit of the FMO complex, as done in most of the previous studies of the energy transfer in the FMO (e.g.\ Refs.~\cite{AmVaGr00__,ReMaKue01_137_,ReMa98_4381_,MiBrGr10_257_,ChFl09_241_,BrStVa05_625_,EnCaRe07_782_,CaChDa09_105106_,PlHu08_113019_,MueMaAd07_16862_,MoReLl08_174106_,AdRe06_2778_,KrKrRo0__,WuLiSh10_105012_,SaChWh11_11906_,BrEi11_051911_}), seems to be a reasonable approximation considering the inter-monomer couplings and the short time scales adopted in the present work.

In the literature there exist many different sets of site energies of the BChls in the FMO (see e.g.~Fig.~3 of Ref.~\cite{OlJaLi11_8609_}).
 Since these sets of energies often differ quite strongly from each other, we have investigated the dependence of the transfer dynamics for two particular sets of energies (taken from Ref.~\cite{OlJaLi11_8609_} and Ref.~\cite{ScMuEl11_93_}) and found large differences.
In particular the amount of excitation arriving on BChl~3 (near the reaction center) may vary by more than 100\% depending on which data set was taken.  
This shows that for further investigations and interpretations of the transport (e.g.\ regarding an optimal efficiency) an accurate determination of the energy landscape within the FMO complex is essential.

Another crucial factor is the initial state, i.e.\ which BChl molecules are excited in the beginning.
We considered different initial conditions, namely when the excitation is localized either on  BChl~1, 6, or~8. This choice of the initial state is motivated by the assumption that BChl~1, 6, or~8 have the highest probability to obtain the excitation from the chlorosome antenna structure. 
Remarkably, when initially only BChl~8 is excited, it was found that a slow exponential-like transfer away from BChl~8 takes place.
This is in contrast to the oscillations found when starting on BChl~1 and the faster transfer away from BChl~1 or~6.
These oscillations have been the subject of study of many research groups, both theoretically and experimentally. 
Ultrafast experiments where BChl 8 is present as well might reveal this strong difference in the energy transfer dynamics away from it.
Furthermore, we found that the influence of variations of different parameters on the transfer strongly depends on the initial state. Here, the parameters we varied include electronic energies, coupling energies between BChl molecules, or the coupling to the environment.
Because of this strong dependence on the initial condition, in further investigations it would be important to get more detailed information about the arrangement of the FMO w.r.t.\ the baseplate and the chlorosomes  and learn more about the transfer of excitation from the chlorosomes to the FMO complex.

Let us briefly compare our findings with those of Refs.~\cite{OlJaLi11_8609_,ScMuEl11_93_}, from which the couplings and energies of the BChls were taken.
In Ref.~\cite{OlJaLi11_8609_} Olbrich et.\ al.\ found an even slower decay of the excitation on BChl~8 than in the present work.
As in the present paper, they also observed a much faster transport away from BChl~1 or~6 compared to BChl~8, but no oscillations, which can be attributed to the much higher (effective) reorganization energy of their spectral density. 
Schmidt am Busch et.\ al.\ found a considerably faster decay of the excitation on BChl~8~\cite{ScMuEl11_93_}.
This can be explained by the fact that in their calculation the strong coupling of BChl~8 to BChl~1 is almost twice as large as the respective coupling in Ref.~\cite{OlJaLi11_8609_}, which was taken in the present work. 
Note that in the calculations of Ref.~\cite{ScMuEl11_93_} the  initial state  is a superposition of excitation on BChls~8, 1, and 2.

We have also investigated various initial states where the excitation is coherently distributed over different BChls (which may be located on different monomers) to investigate the possibility of phase directed transport \cite{Ei11_33_}.
For all cases considered the initial phase had a marginal effect on the transport dynamics.
However the initial population strongly influences the transport dynamics.
Thus in further studies to avoid the arbitrary choice of the initial state and to take the flow off of excitation into account, also the chlorosome (at least the part close to the FMO) and the reaction center (which plays a significant role) should be included into a holistic simulation. 
Efforts towards this goal are being made in our research groups.

\acknowledgement
Financial support from the DFG under Contract No. Ei 872/1-1 is acknowledged.
We thank John Briggs for helpful comments and Thomas Renger and Ulrich Kleinekath\"ofer for interesting discussions. 



\providecommand*\mcitethebibliography{\thebibliography}
\csname @ifundefined\endcsname{endmcitethebibliography}
  {\let\endmcitethebibliography\endthebibliography}{}

\begin{table}[b]
\begin{tabular}{ c| c c c c c c c c }
   &       2   &       3   &       4   &       5   &       6   &       7   &       8   \\ 
\hline
1  &\bf{-80.3} &      3.5  &     -4.0  &      4.5  &\bf{-10.2} &     -4.9  &\bf{ 21.0} \\
2  &           &\bf{ 23.5} &      6.7  &      0.5  &      7.5  &      1.5  &      3.3  \\
3  &           &           &\bf{-49.8} &     -1.5  &     -6.5  &      1.2  &      0.7  \\
4  &           &           &           &\bf{-63.4} &\bf{-13.3} &\bf{-42.2} &     -1.2  \\
5  &           &           &           &           &\bf{ 55.8} &      4.7  &      2.8  \\
6  &           &           &           &           &           &\bf{ 33.0} &     -7.3  \\
7  &           &           &           &           &           &           &     -8.7  \\
\end{tabular}
\caption{Intra-monomer couplings from Ref.~\cite{OlJaLi11_8609_} for the trimer structure in $\cm$, obtained using MD simulations.
Values greater than $10\cm$ are highlighted in bold face.}
\label{tab:coup_monomer}
\end{table}

\begin{table}[b]
\begin{tabular}{ c| c c c c c c c c }
   &   1  &   2  &   3  &   4  &   5  &   6  &   7  &   8  \\
\hline
1  &  1.0 &  0.3 & -0.6 &  0.7 &  2.3 &  1.5 &  0.9 &  0.1 \\
2  &  1.5 & -0.4 & -2.5 & -1.5 &  7.4 &  5.2 &  1.5 &  0.7 \\
3  &  1.4 &  0.1 & -2.7 &  5.7 &  4.6 &  2.3 &  4.0 &  0.8 \\
4  &  0.3 &  0.5 &  0.7 &  1.9 & -0.6 & -0.4 &  1.9 & -0.8 \\
5  &  0.7 &  0.9 &  1.1 & -0.1 &  1.8 &  0.1 & -0.7 &  1.3 \\
6  &  0.1 &  0.7 &  0.8 &  1.4 & -1.4 & -1.5 &  1.6 & -1.0 \\
7  &  0.3 &  0.2 & -0.7 &  4.8 & -1.6 &  0.1 &  5.7 & -2.3 \\
8  &  0.1 &  0.6 &  1.5 & -1.1 &  4.0 & -3.1 & -5.2 &  3.6 \\
\end{tabular}
\caption{Inter-monomer couplings from Ref.~\cite{OlJaLi11_8609_}  for the trimer structure in $\cm$, obtained using MD simulations.
Upper triangle: couplings between monomer units A-B, B-C and C-A. Lower triangle: couplings between monomer units A-C, B-A and C-B.}
\label{tab:coup_trimer}
\end{table}

\begin{table}[b]
\begin{tabular}{ l| c c c c c c c c }
 & 1 & 2 & 3 & 4 & 5 & 6 & 7 & 8 \\ 
\hline
Olbrich et al.~\cite{OlJaLi11_8609_} & 186 & 81 &  0 & 113 & 65 & 89 & 492 & 218 \\
Schmidt am Busch et al.~\cite{ScMuEl11_93_} & 310 & 230 & 0 & 180 & 405 & 320 & 270 & 505 \\
\end{tabular}
\caption{\label{tab:site_energies}Site energies of BChl 1--8 on each monomer.
The values are given in $\cm$.
The respective lowest value is set to zero (for the calculation of the transfer only the energy differences are relevant).
Note that for the values of Olbrich  et al.\ the average positions from Table 1 of Ref.~\cite{OlJaLi11_8609_} are taken.
}
\end{table}


\end{document}